\input{aipcheck}

\documentclass[
    ,final            
  ]
  {aipproc}

\layoutstyle{6x9}

\newcommand {\hi}       {{\sc H~i}}
\newcommand {\kms}	{{\mbox{$\rm \,km/s$}}}
\newcommand {\arcmin}	{{\mbox{$'$}}}

\begin{document}

\title{Mapping Hydrogen in the Galaxy, Galactic Halo, and Local Group with ALFA:\\The GALFA - \hi\ Survey Starting with TOGS}

\classification{
98.35.Gi, 	
98.38.Am, 	
98.38.Dq, 	
98.38.Gt, 	
98.56.Ne, 	
98.56.Tj 	
}

\keywords      {
Arecibo Telescope,
\hi\ 21cm-line surveys,
Galactic interstellar medium,
cold diffuse clouds,
magnetic fields,
Galactic halo,
high-velocity clouds,
Magellanic Stream,
M33
}

\author{S. J. Gibson}{
  address={
	Arecibo Observatory, 
	HC 3 Box 53995, Arecibo, PR 00612}
}

\author{K. A. Douglas}{
  address={
	Space Sciences Laboratory, University of California, Berkeley,
	CA 94720}
}

\author{C. Heiles}{
  address={
	Department of Astronomy, University of California, Berkeley, CA 94720}
}

\author{E. J. Korpela}{
  address={
	Space Sciences Laboratory, University of California, Berkeley,
	CA 94720}
}

\author{J. E. G. Peek}{
  address={
	Department of Astronomy, University of California, Berkeley, CA 94720}
}

\author{M. E. Putman}{
  address={
	Department of Astronomy, University of Michigan, Ann Arbor, MI 48109}
}

\author{S. Stanimirovi\'{c}}{
  address={
	Department of Astronomy, University of Wisconsin, Madison, WI 53706}
}

\begin{abstract}

Radio observations of gas in the Milky Way and Local Group are vital for
understanding how galaxies function as systems.  The unique sensitivity of
Arecibo's 305m dish, coupled with the 7-beam Arecibo L-Band Feed Array (ALFA),
provides an unparalleled tool for investigating the full range of interstellar
phenomena traced by the \hi\ 21cm line.
The GALFA (Galactic ALFA) \hi\ Survey is mapping the entire Arecibo sky over a
velocity range of $-700$ to $+700$~\kms\ with 0.2~\kms\ velocity channels and
an angular resolution of 3.4\arcmin.  We present highlights from the TOGS (Turn
on GALFA Survey) portion of GALFA - \hi, which is covering thousands of square
degrees in commensal drift scan observations with the ALFALFA and AGES
extragalactic ALFA surveys.
This work is supported in part by the National Astronomy and Ionosphere Center,
operated by Cornell University under cooperative agreement with the National
Science Foundation.

\end{abstract}

\maketitle




\begin{figure}
  \includegraphics[height=.3\textheight]{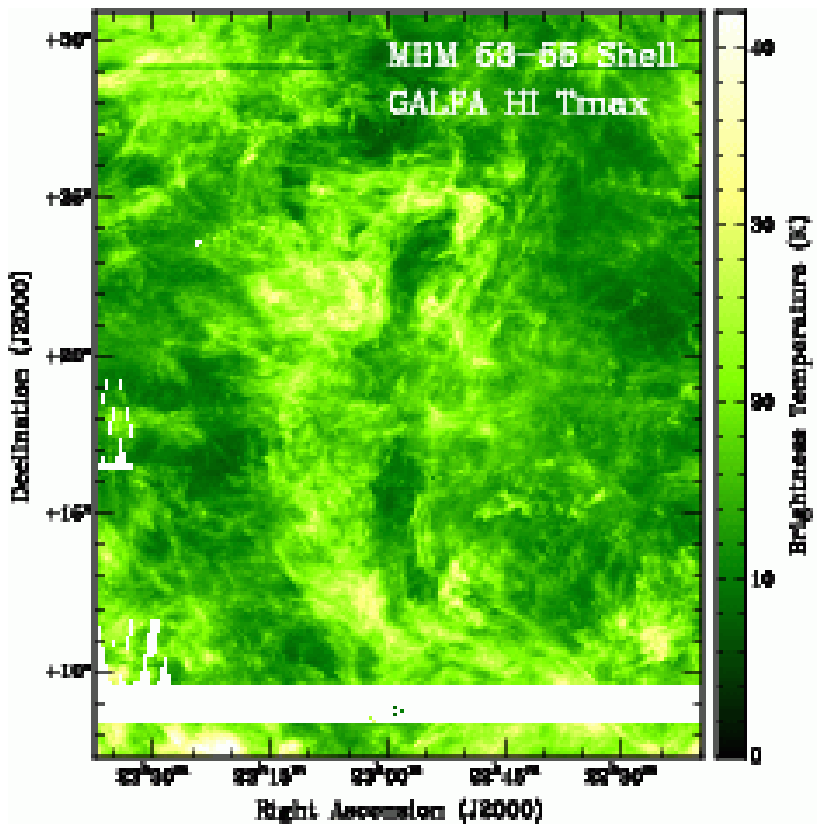}
  \includegraphics[height=.3\textheight]{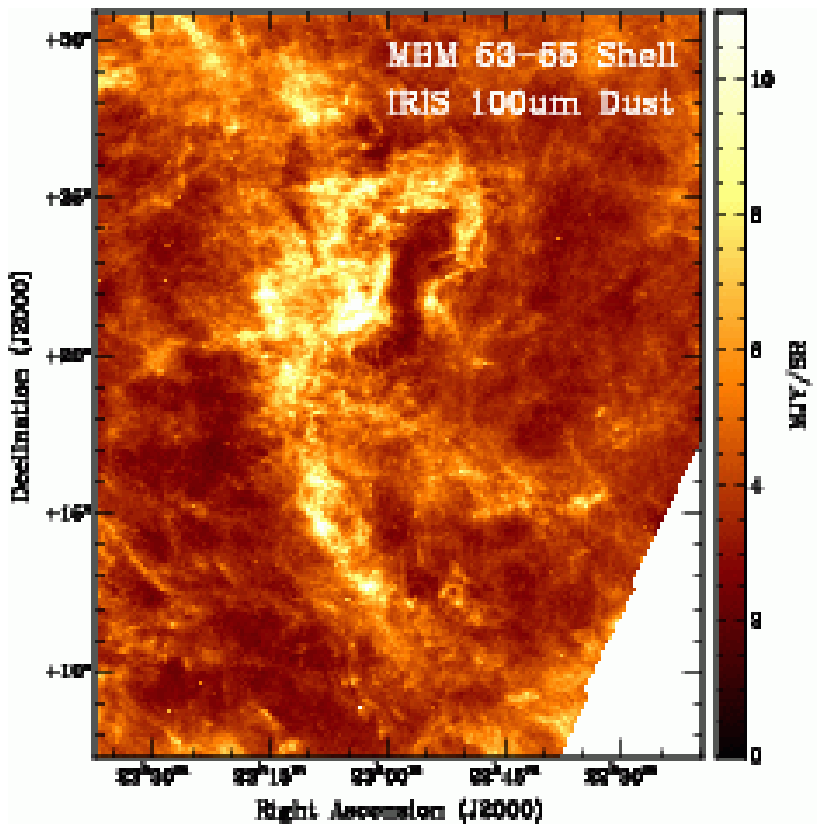}
  \caption{
{\it (LEFT)\/} GALFA map of \hi\ line peak brightness and {\it (RIGHT)\/} IRAS
100 micron thermal dust emission from the nearby molecular cloud complex MBM
53-55 \cite{mbm85}, which has not yet formed many stars, but is likely to do so
in the future \cite{y03}.  The complex appears to be part of a large dynamic
structure, perhaps an expanding shell \cite{g94}.  GALFA sensitivity and
resolution allows every IRAS filament to be velocity-mapped in \hi, so the
kinematics and temperature structure of the entire region can be thoroughly
investigated.  Many narrow-line \hi\ features are present, indicating
widespread cold atomic gas from which the molecular clouds may still be
condensing (Gibson et al., in prep).
}
\end{figure}

\begin{figure}
  \includegraphics[height=.2\textheight]{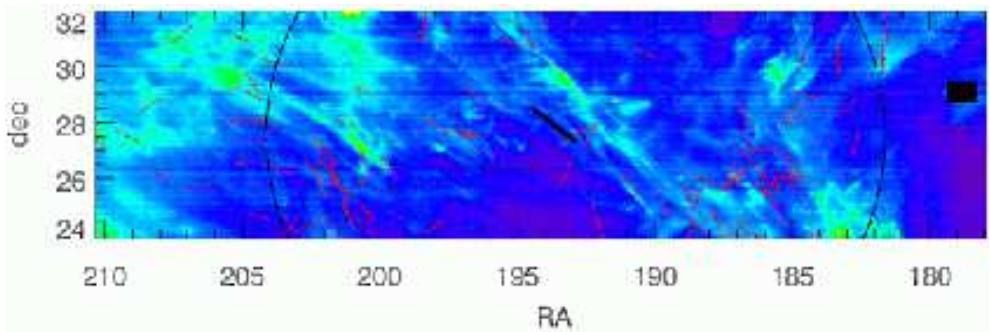}
  \caption{
Bundles of very fine ($\sim 5$\arcmin\ wide) filamentary \hi\ features near the
north Galactic pole.  These features are highly correlated with optical
starlight polarization measurements (red lines), implying magnetic fields
running along their lengths.  They are also roughly parallel to the local
spiral arm.  We find these ``filament bundles'' throughout the diffuse
interstellar medium, suggesting a new technique for finding the orientation of
magnetic fields (Peek et al, in prep).
}
\end{figure}

\begin{figure}
  \includegraphics[height=.35\textheight]{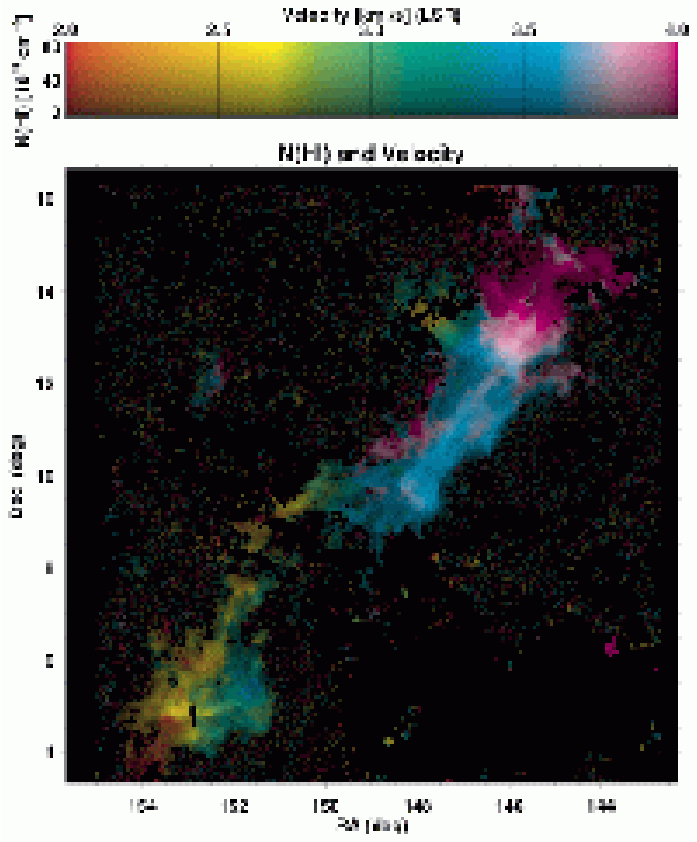}
  \includegraphics[height=.35\textheight]{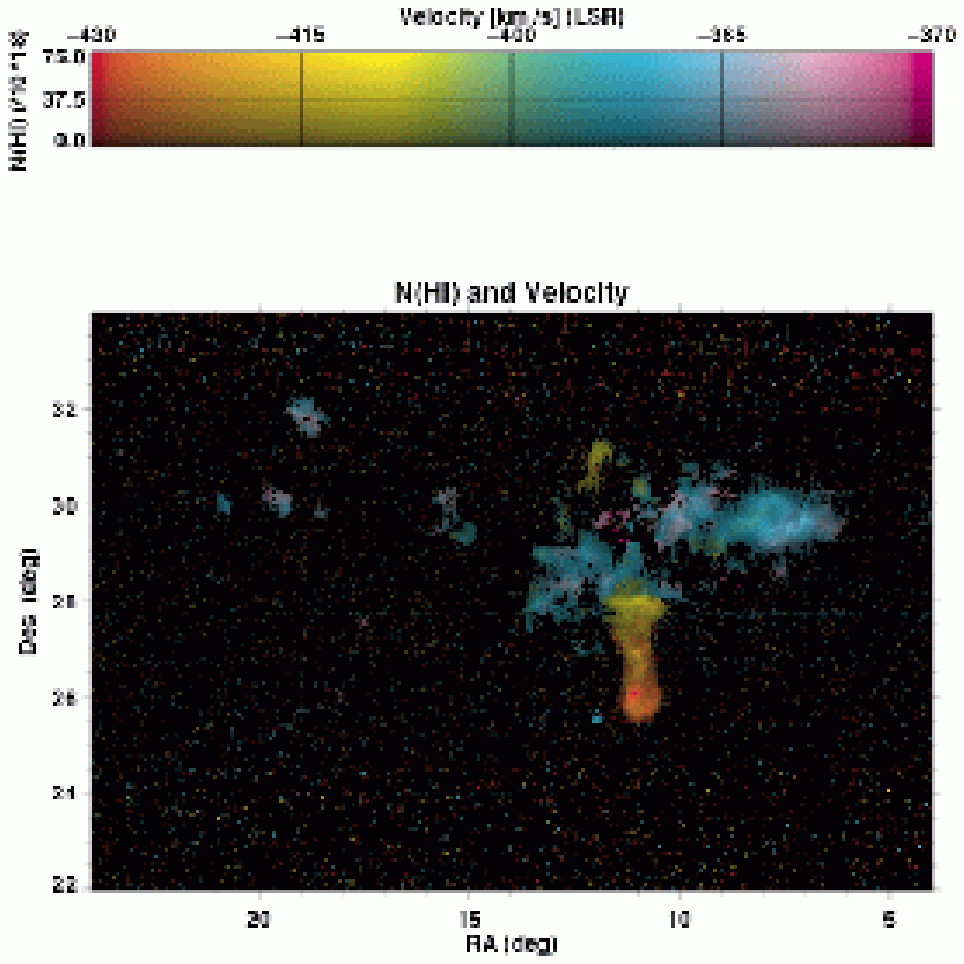}
  \caption{
{\it (LEFT)\/} GALFA image of an extremely cold (17 K) and nearby (< 40 pc)
\hi\ cloud \cite{m06}.  Color represents velocity along the sight line, and 
brightness represents hydrogen column density.  Because of its location within
the local hot bubble, this cloud makes a very interesting laboratory for the
study of cold \hi\ and cold/hot gas interfaces. A comparison of the 21-cm
observations to infrared dust emission and optical and UV stellar absorption is
underway (Peek et al., in prep).
{\it (RIGHT)\/} GALFA reveals detailed structure of Very High-Velocity Clouds
(VHVCs) in the halo and can be used to probe the diffuse halo medium through
fingers extending off the sides of the cloud and head-tail clouds
(e.g. \cite{p07,s06}; Peek et al., in prep; Grcevich et al., in prep).
}
\end{figure}

\begin{figure}
  \includegraphics[height=.25\textheight]{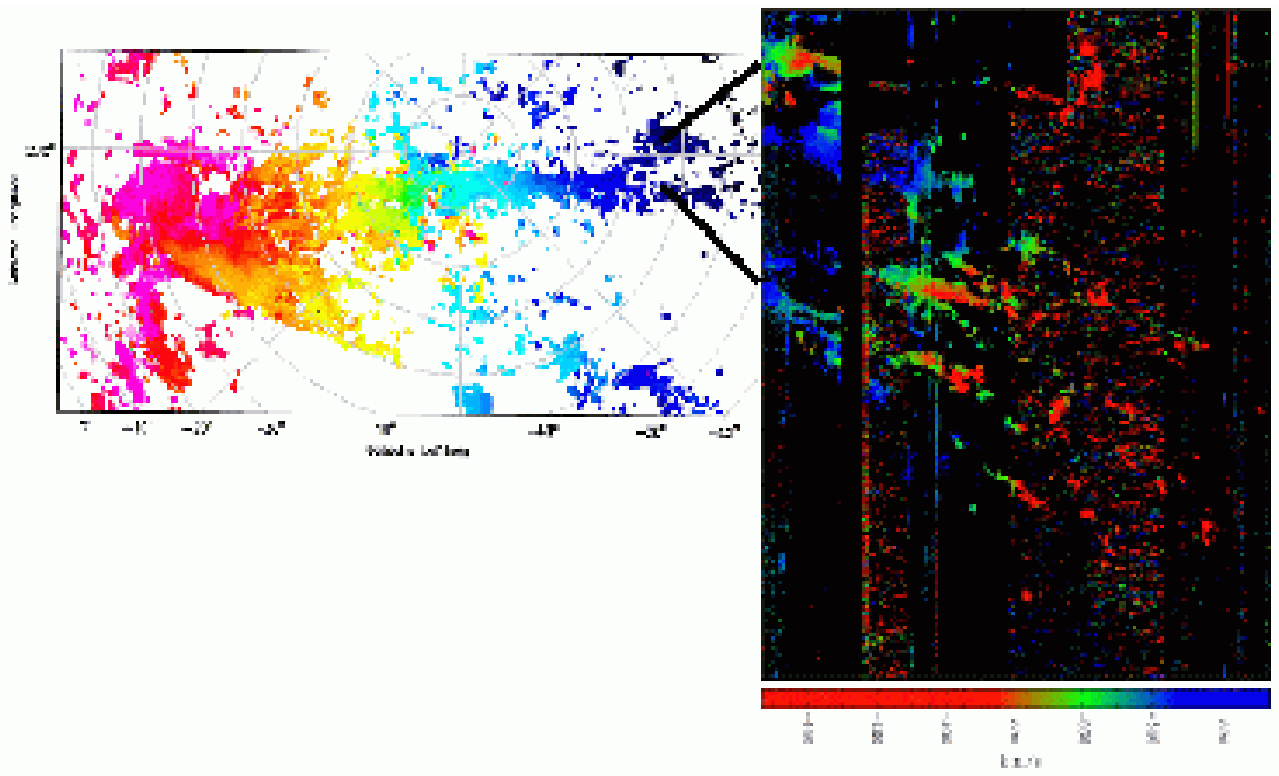} 
  \includegraphics[height=.25\textheight]{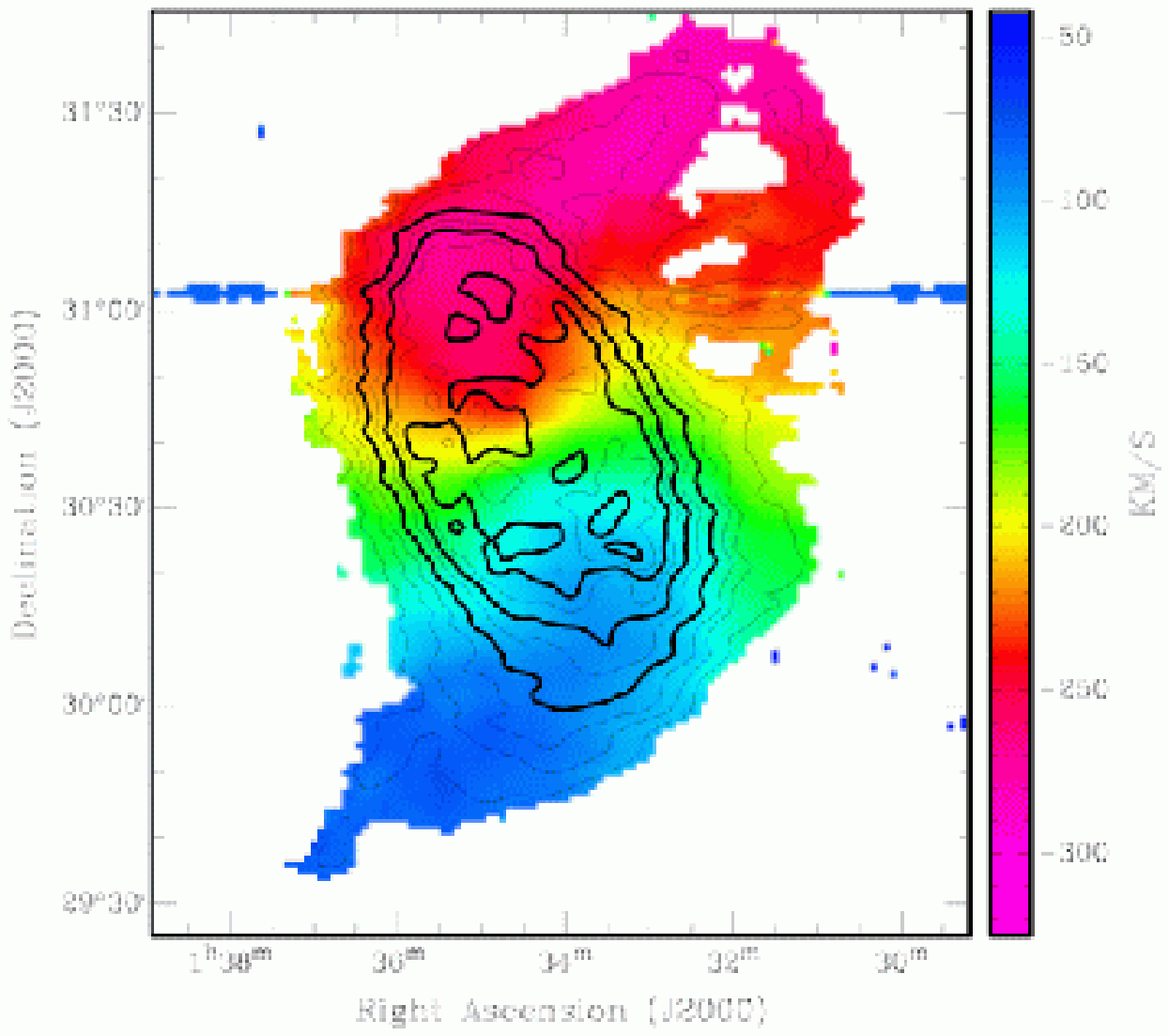}
  \caption{ 
{\it (LEFT)\/} The \hi\ velocity field of the Magellanic System from the Parkes
telescope (15.5\arcmin\ resolution; \cite{p03}).  {\it (CENTER)\/} the tip of
the Magellanic Stream as observed by GALFA, revealing four coherent 10-15
degree long filaments that differ in morphology and velocity structure and may
have different origins/ages.  Numerous small clouds with high negative
velocities are also evident.  Some of these clouds show evidence for a
multiphase medium and may result from spatial fragmentation of the Stream due
to thermal instability (Stanimirovic et al., in prep).
{\it (RIGHT)\/} The Local Group galaxy M33 as seen by GALFA.  What was
traditionally thought to be a quiescent galaxy shows clear evidence for 
tidal/ram pressure disruption and/or gas cloud accretion (Putman et al., in
prep).
}
\end{figure}





\bibliographystyle{aipprocl} 

%


\end{document}